\documentclass{article}
\usepackage{spconf,amsmath,amsfonts,float,graphicx,hyperref,comment,multirow}
\usepackage{caption, subcaption}
\usepackage[table,xcdraw]{xcolor}

\def\x{{\mathbf x}}
\def\t{{\mathbf t}}
\def\y{{\mathbf y}}
\def\f{{\mathbf f}}

\title{Authentication of copy detection patterns under machine learning attacks: a supervised approach}
\name{B. Pulfer, R. Chaban, Y. Belousov, J. Tutt, O. Taran, T. Holotyak, S. Voloshynovskiy\thanks{S. Voloshynovskiy is a corresponding author.
This research was partially funded by the Swiss National Science Foundation SNF No. 200021 182063.}}
\address{University of Geneva\\
Department of Computer Science\\
Switzerland}

\begin{document}
%\ninept
%
\maketitle
\begin{abstract}
Copy detection patterns (CDP) are an attractive technology that allows manufacturers to defend their products against counterfeiting. The main assumption behind the protection mechanism of CDP is that these codes printed with the smallest symbol size (1x1) on an industrial printer cannot be copied or cloned with sufficient accuracy due to data processing inequality. However, previous works have shown that Machine Learning (ML) based attacks can produce high-quality fakes, resulting in decreased accuracy of authentication based on traditional feature-based authentication systems. While Deep Learning (DL) can be used as a part of the authentication system, to the best of our knowledge, none of the previous works has studied the performance of a DL-based authentication system against ML-based attacks on CDP with 1x1 symbol size. In this work, we study such a performance assuming a supervised learning (SL) setting.
\end{abstract}

\section{Introduction}
\label{sec:intro}
Counterfeited products negatively affect world economy. They are present in multiple industries, varying from pharmaceutical and luxury products to identification documents, banknotes and even food and agricultural products.

Among all available technologies to defend against counterfeiting, copy detection patterns (CDP), based on hand-crafted randomness, represent a convenient, efficient and user-friendly solution \cite{picard2004, picard2021acm}. The advancement of mobile phone cameras' resolutions empowers the CDP technology even further, allowing the verification of authenticity of a product by the end customers. Furthermore, CDP are easily integrable into product design, are applicable in many different areas and feature low computational complexity for enrollment and authentication.

Previous works have shown that CDP with a symbol size of 3x3 and 5x5 printed on desktop printers at 600 dpi might be cloned under certain conditions \cite{taran2021reliable, Taran2019icassp, taran2021mobile}. More recent works, which used industrial printers such as HP Indigo capable of producing symbols of size 1x1 at the resolution of 812.8 dpi, have shown that attacks based on ML-methods make the authentication based on hand-crafted features and SVM-based classifiers harder for the defender \cite{chaban2021machine, yadav:hal-02330988}.

\begin{figure}[t]
    \centering
    \includegraphics[width=1\linewidth]{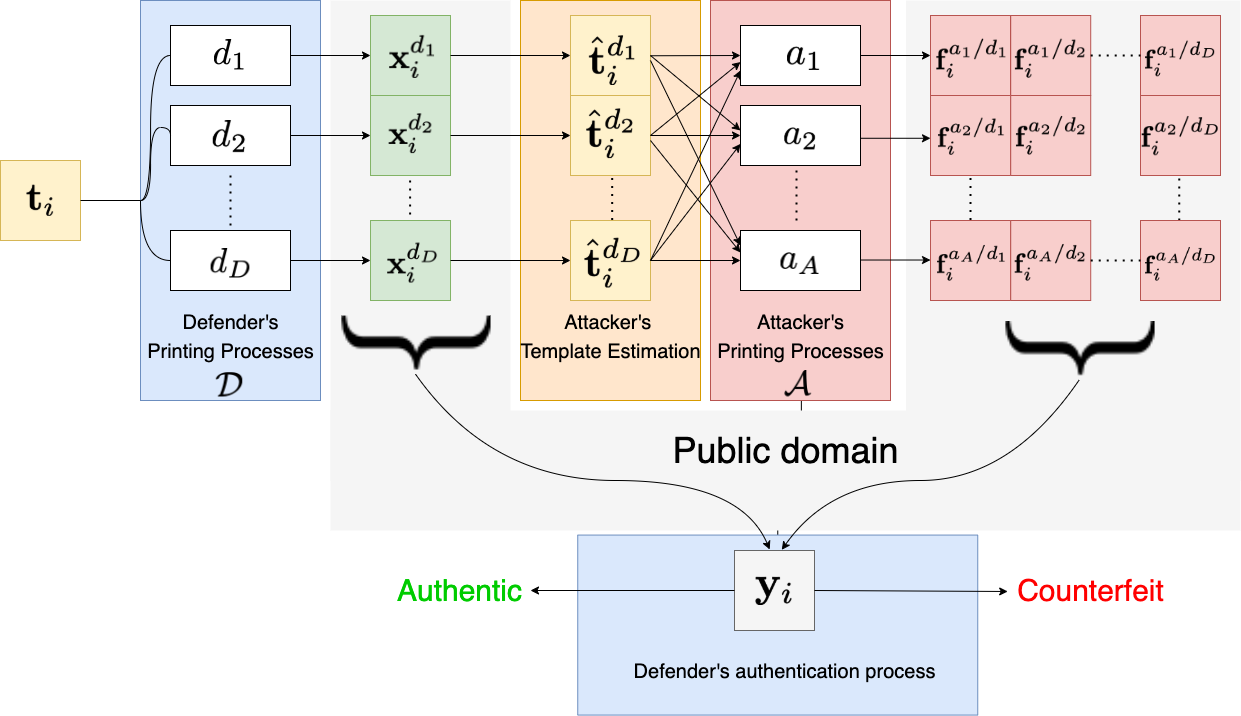}
    \caption{Schematic representation of the problem. The defender creates a digital code $\t_i$, which is printed using the set of original printing processes $\mathcal{D}$. The printed original CDP are made public. An attacker estimates the digital template $\t_i$ given an original code $\x_i^d$ as $\hat{\t}_i^d$, and prints the estimations through its printing processes $\mathcal{A}$. The defender has to design an authentication system that, given original or fake code $\y_i$, predicts its authenticity with respect to $\t_i$.}
    \label{fig:problem}
\end{figure}

However, up to our best knowledge, none of the previous works has combined CDP printed on industrial printers, and attacked using ML-based attacks, with Deep Learning (DL) based authentication strategies. To fill this gap, in this study we empirically evaluate the capabilities of a DL-based authentication system to detect ML-based counterfeits. 

We assume a supervised learning (SL) setting, where the defender possesses of at least one type of non-authentic codes. In our case, this assumption translates in the defender observing fakes produced by the attacker using one (potentially more) particular printing process(es). We refer to "distribution shift" of fakes for cases where the
fake codes used during the training and test are printed on different printers. We address the following research questions (RQ):

\textbf{RQ1: Is it possible to reliably detect ML-based fakes using a two-class classifier?}
We measure the performance of the supervised classifier w.r.t. a test set of original and fake codes printed on the same printer as the training data.

\textbf{RQ2: Is the supervised classifier robust to a distribution shift of fakes?}
We measure the performance of our system against fake CDP, which were printed on a different printer w.r.t. CDP used for training. We are interested in learning if the system is robust to such a distribution shift. Also, we investigate which type of fake codes should the defender use during training such that the authentication system yields the best performance under a distribution shift.

\section{Problem formulation}
\label{sec:problem}
To defend its products from counterfeiting, a manufacturer (or defender) creates $N$ digital templates $\t_i \in \{0, 1\}^{w \times h}, i \in \mathcal{I} = \{1, ..., N\}$ of size $w \times h$ that are then printed through one or more defender's printing processes $d \in \mathcal{D}$  ($|\mathcal{D}| = D$), to obtain original CDP $\x_i^{d} \in \mathbb{R}^{w \times h}$. The original code $\x_i^{d}$ will be available to the customer upon purchase, which in turn can verify the authenticity of the product by acquiring it.

An attacker accesses the original printed CDP in the public domain and estimates, for an observed original $\x_i^{d}$, the digital template $\t_i$ as $\mathbf{\hat{t}}_i^{d}$ through a ML-based estimation model \cite{Taran2019icassp, taran2021mobile, chaban2021machine, yadav:hal-02330988}. Having the estimation $\mathbf{\hat{t}}_i^{d}$, the attacker is then able to produce a fake code $\f_i^{a/d} \in \mathbb{R}^{w \times h}$ using one of the attacker's printing processes $a \in \mathcal{A}, |\mathcal{A}|=A$. The superscript "$a/d$" denotes the fact that the digital template $\mathbf{\hat{t}}^d_i$ is estimated from the CDP $\x_i^d$ printed on printer $d$ by the defender and reproduced on the printer $a$ by the attacker. The resulting fake is then put into the public domain as shown in \autoref{fig:problem}. The attacker can thus generate $D \times A$ fake CDP for a single template. We denote $\x^d=\{\x^d_i| i \in \mathcal{I}\}$ and $\f^{a/d}=\{\f^{a/d}_i| i \in \mathcal{I}\}$.

The goal of the defender is to design an authentication system which, given a probe $\y_i \in \{\x_i^{d}, \f_i^{a/d} | d \in \mathcal{D} \land a \in \mathcal{A} \}$, can correctly determine whether the probe is an original one, i.e., $\y_i \in \{\x_i^{d} | d \in \mathcal{D}\}$ or fake, i.e., $\y_i \in \{\f_i^{a/d} | d \in \mathcal{D} \land a \in \mathcal{A}\}$. The authentication system should minimize the probability of accepting fake codes, $P_{fa}$, while also minimizing the probability of missing original codes, $P_{miss}$. Notice that, because of the phenomenon of \textit{dot gain}, which is related to the interaction
between the printing ink and the substrate, it is assumed that it is generally not possible for the attacker to generate fakes that are identical to the originals \cite{picard2004, picard2021acm}. 

In this study, we assume that the attacker has an access to the same printing processes as the defender, that is: $\mathcal{D} \subseteq \mathcal{A}$. This represents the worste case for the defender, as the difference in printing artifacts is minimal. We will validate this assumption in \autoref{sec:results}. Furthermore, while it would be possible for the manufacturer to use the set of enrolled printed codes $\{\x_i^{d}| d \in \mathcal{D}\}$ to help authenticate a probe $\y_i$, we assume that the defender does not use physical references and relies only on the set of digital templates $\{\t_i| i \in \mathcal{I}\}$ when authenticating probe $\y_i$. This assumption is due to a typical operational industrial scenario where the defender tries to minimize its costs for the enrollment of physical references for each printed item and Information Technology (IT) infrastructure management. Thus, the authentication is performed based on the digital reference only. Finally, we focus on supervised learning based authentication systems. We thus assume that the defender has an access to at least one (or more) set(s) of fake CDP $\f^{a/d}$, which could have been observed from the public domain, for the authentication system training.

\begin{figure}[t]
    \centering
    \includegraphics[width=1\linewidth]{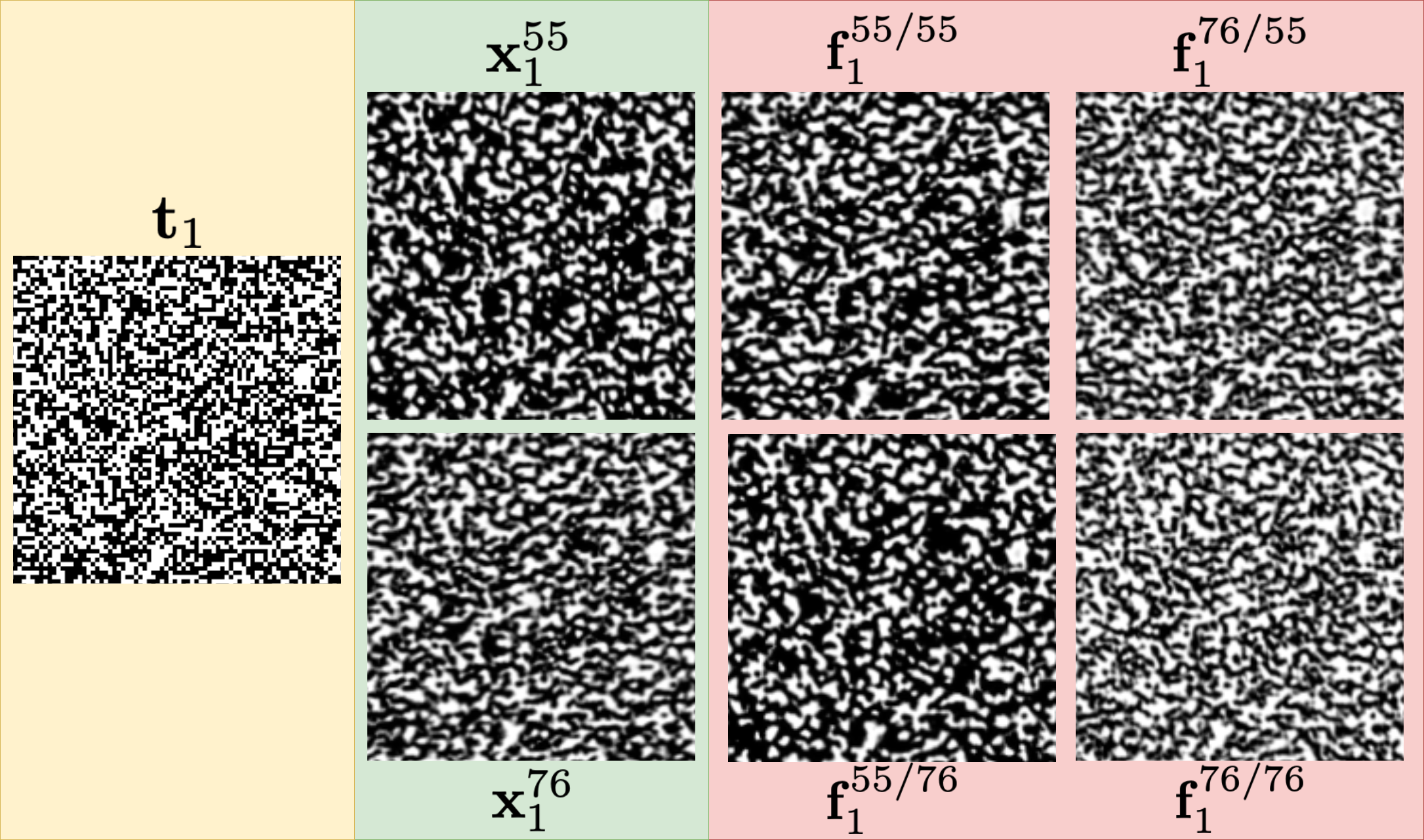}
    \caption{Example of the first tuple of our dataset. For a single template code (cropped part), two originals are printed on two industrial printers. Fakes are obtained by estimating the digital template starting from the two original printed CDP and then printing the estimations on both printers.}
    \label{fig:tuple}
\end{figure}

\section{Dataset}
\label{sec:data}
\begin{figure}[t]
    \centering
    \begin{subfigure}[b]{0.237\textwidth}
        \centering
        \includegraphics[width=\textwidth]{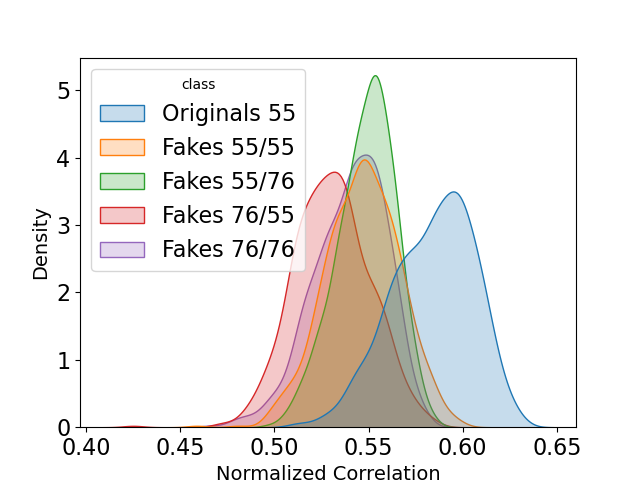}
        \caption{$\x^{55}$ and fakes}
        \label{fig:kde55}
    \end{subfigure}
    \hfill
    \begin{subfigure}[b]{0.237\textwidth}
        \centering
        \includegraphics[width=\textwidth]{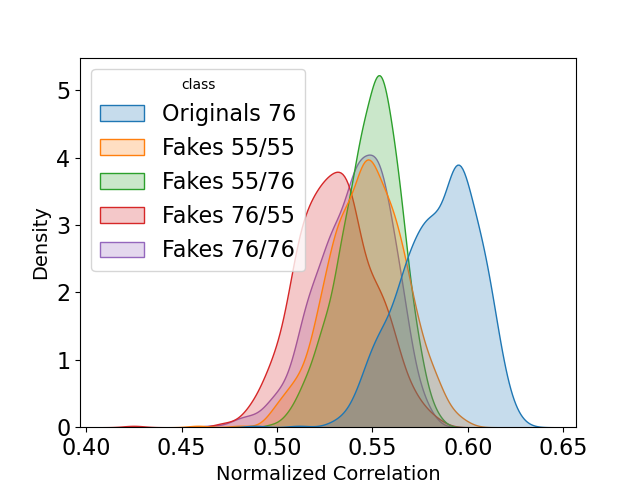}
        \caption{$\x^{76}$ and fakes}
        \label{fig:kde76}
    \end{subfigure}
    \caption{KDE plot of normalized correlation between the digital templates and printed original and fake codes.}
    \label{fig:kde}
\end{figure}

We use 720 tuples of CDP of size 1x1 pixel from the Indigo 1x1 base dataset
presented in \cite{chaban2021machine} and publicly available at \cite{dataset}, where each tuple includes one digital template, two original and four fake printed codes as shown in \autoref{fig:tuple}. Each digital template is a binary image of size 684x684 pixels, whereas printed originals and fakes are gray-scale images of the same size obtained after synchronization based on special markers. Templates were printed using industrial printers HP Indigo 5500 (HPI55) and HP Indigo 7600 (HPI76) with resolution 812.8 dpi. Originals are 2400 ppi scans of the obtained printed templates (to a simulate mobile phone camera), while for fakes 6400 ppi scans were used to obtain the ML-estimations that were then printed with the same conditions \cite{chaban2021machine}. We denote the printing processes based on the printer number as $\{55, 76\}$. Tuple $i$ of our dataset is thus presented as $(\t_i, \x^{55}_i, \x^{76}_i, \f^{55/55}_i, \f^{55/76}_i, \f^{76/55}_i, \f^{76/76}_i)$, where $\t$, $\x$ and $\f$ represent digital templates, original and fake printed codes respectively. The superscript for originals indicates the used printer. For fakes, the superscript $a/d$ indicates that the estimation of the template was originated from original $\x_i^d$ and then printed with printer $a$. In \autoref{fig:kde}, we show the Kernel Density Estimation (KDE) plot of normalized correlation with the template codes for both original and all fake. The plots show that while originals (shown in blue) correlate more with templates, there is a decent overlap with fakes, making the authentication task challenging. Therefore the linear separation of originals and fakes in the considered metric is not feasible.

\section{Methodology}
\label{sec:method}

\begin{figure}[t]
    \centering
    \includegraphics[width=0.8\linewidth]{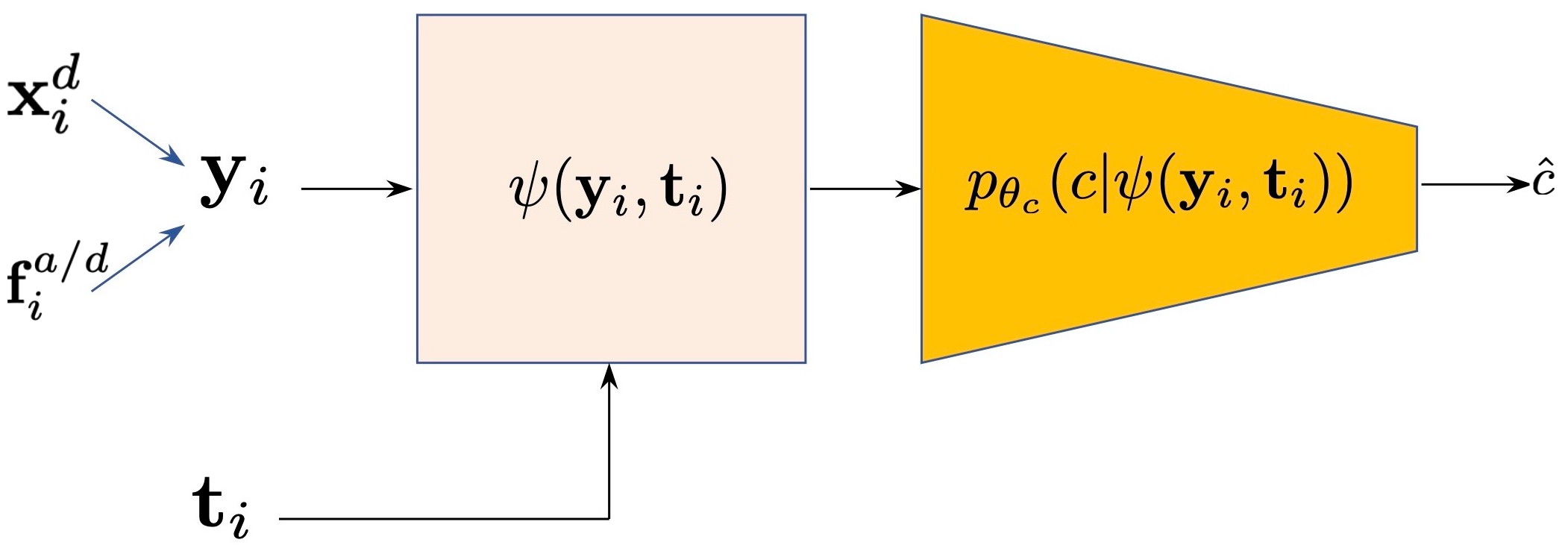}
    \caption{Considered system of supervised authentication based on reference digital template $\t_i$. The sample under investigation $\y_i$, which could be an original $\x_i^d$ or fake $\f_i^{a/d}$, is aggregated with the corresponding template $\t_i$ through aggregating function $\psi(.)$. The deep classifier $p_{\theta_c}$, parametrized by learnable parameters $\theta_c$, outputs binary prediction $p_{\theta_c}(c|\psi(\y_i, \t_i)) = \hat{c}$ indicating the degree of acceptance of $\y_i$ as an original code.}
    \label{fig:method}
\end{figure}

The proposed supervised model takes as an input an aggregation $\psi(\y_i, \t_i)$ of the code under investigation $\y_i$ with the respective template $\t_i$. In the general case, the aggregation might be performed in different ways, for example, channel-wise concatenation, subtraction, and so on. In our experiments, $\psi(.)$ is the channel-wise concatenation, such that $\psi(\y_i, \t_i) \in\mathbb{R}^{2 \times 684 \times 684}$. The classifier $p_{\theta_c}(c|\psi(\y_i, \t_i))$, parametrized by learnable parameters $\theta_c$, is trained to output $c=1$ when $\y_i$ is an original CDP, and $c=0$ when it is a counterfeit. The investigated system architecture is schematically shown in \autoref{fig:method}. Let us denote $\phi_{\theta_c}(\y, \t) = p_{\theta_c}(c | \psi(\y, \t))$ for simplicity. Our model is trained to minimize the following loss function:

\begin{equation}
\mathcal{L}(\theta_c) = \frac{1}{N}\sum_{i=1}^{N} \Bigg[\mathcal{L}_{miss}(i) + \mathcal{L}_{fa}(i) \Bigg],
\end{equation}

where:

\begin{equation}
    \mathcal{L}_{miss}(i) =  \frac{1}{D}\sum_{d \in D}d_{BCE}(\phi_{\theta_c}(\x_i^d, \t_i), 1),
\end{equation}

\begin{equation}
    \mathcal{L}_{fa}(i) = \frac{1}{AD} \sum_{a \in \mathcal{A}, d \in \mathcal{D}}d_{BCE}(\phi_{\theta_c}(\f_i^{a/d}, \t_i), 0),
\end{equation}

and the binary cross-entropy loss is defined as:

\begin{equation}
d_{BCE}(\hat{y},y) = - y\log(\hat{y}) - (1-y)\log(1-\hat{y}).
\end{equation}

\subsection{Setups}
\label{subsec:setups}
We conduct experiments on setups which represent interesting study cases for the defender. \textbf{Setup 1:} The defender produces originals and observes fakes from a single printer. \textbf{Setup 2:} The defender prints on one printer but observes fakes from multiple printers. \textbf{Setup 3:} The defender prints multiple originals and observers multiple types of fakes.

The first setup could represent, for example, a situation where a defender prints originals with a unique printing process and produces his own fakes with some kind of manipulation (e.g., scanning original and re-printing).
The second setup is as the first one, except that the defender has a way to create a multitude of manipulations to its original codes.
Finally, in the third setup the defender also possesses multiple printing processes (e.g., multiple printers).

\subsection{Training}
For each setup, we train a standard ResNet18 backbone \cite{he2015deep} with an added 2-layer MLP head. Each model was trained with a 40/10/50\% train-validation-test split for 1'000 epochs using early stopping and a learning rate of $0.005$ \footnote{The Github repository with a source code will be available upon paper acceptance.}. 

For training data, we use the following random augmentations: horizontal flip, vertical flip, $90^\circ$ rotations, and gamma correction with $\gamma \in [0.4, 1.3]$. Each augmentation is chosen randomly, but then applied equally to all CDP in the tuple. For example, a clockwise rotation of $90^\circ$ is applied to the template $\t_i$ as well as to all $\x_i^d$ and $\f_i^{a/d}$ used in the specific setup. No augmentation is carried out for validation and test data.

\section{Results}
\label{sec:results}

\begin{table}[t]
\begin{tabular}{|ccccccc|}
\hline
\multicolumn{1}{|c|}{$\mathcal{D}_{train}$} &
  \multicolumn{1}{c|}{$\mathcal{A}_{train}$} &
  \multicolumn{1}{c|}{$\mathcal{D}_{test}$} &
  \multicolumn{1}{c|}{$\mathcal{A}_{test}$} &
  \multicolumn{1}{c|}{$P_{miss}$} &
  \multicolumn{1}{c|}{$P_{fa}$} &
  AUC \\ \hline
\multicolumn{7}{|c|}{\textbf{Setup 1}} \\ \hline
\multicolumn{1}{|c|}{\multirow{2}{*}{$\x^{55}$}} &
  \multicolumn{1}{c|}{\multirow{2}{*}{$\f^{55/55}$}} &
  \multicolumn{1}{c|}{\multirow{2}{*}{$\x^{55}$}} &
  \multicolumn{1}{c|}{$\f^{55/55}$} &
  \multicolumn{1}{c|}{0.01} &
  \multicolumn{1}{c|}{0.00} &
  \textbf{1.00} \\ \cline{4-7} 
\multicolumn{1}{|c|}{} &
  \multicolumn{1}{c|}{} &
  \multicolumn{1}{c|}{} &
  \multicolumn{1}{c|}{$\f^{76/55}$} &
  \multicolumn{1}{c|}{0.01} &
  \multicolumn{1}{c|}{0.00} &
  \textbf{1.00} \\ \hline
\multicolumn{1}{|c|}{\multirow{2}{*}{$\x^{55}$}} &
  \multicolumn{1}{c|}{\multirow{2}{*}{$\f^{76/55}$}} &
  \multicolumn{1}{c|}{\multirow{2}{*}{$\x^{55}$}} &
  \multicolumn{1}{c|}{$\f^{55/55}$} &
  \multicolumn{1}{c|}{0.00} &
  \multicolumn{1}{c|}{1.00} &
  0.70 \\ \cline{4-7} 
\multicolumn{1}{|c|}{} &
  \multicolumn{1}{c|}{} &
  \multicolumn{1}{c|}{} &
  \multicolumn{1}{c|}{$\f^{76/55}$} &
  \multicolumn{1}{c|}{0.00} &
  \multicolumn{1}{c|}{0.00} &
  \textbf{1.00} \\ \hline
\multicolumn{1}{|c|}{\multirow{2}{*}{$\x^{76}$}} &
  \multicolumn{1}{c|}{\multirow{2}{*}{$\f^{55/76}$}} &
  \multicolumn{1}{c|}{\multirow{2}{*}{$\x^{76}$}} &
  \multicolumn{1}{c|}{$\f^{55/76}$} &
  \multicolumn{1}{c|}{0.01} &
  \multicolumn{1}{c|}{0.00} &
  \textbf{1.00} \\ \cline{4-7} 
\multicolumn{1}{|c|}{} &
  \multicolumn{1}{c|}{} &
  \multicolumn{1}{c|}{} &
  \multicolumn{1}{c|}{$\f^{76/76}$} &
  \multicolumn{1}{c|}{0.01} &
  \multicolumn{1}{c|}{1.00} &
  0.26 \\ \hline
\multicolumn{1}{|c|}{\multirow{2}{*}{$\x^{76}$}} &
  \multicolumn{1}{c|}{\multirow{2}{*}{$\f^{76/76}$}} &
  \multicolumn{1}{c|}{\multirow{2}{*}{$\x^{76}$}} &
  \multicolumn{1}{c|}{$\f^{55/76}$} &
  \multicolumn{1}{c|}{0.00} &
  \multicolumn{1}{c|}{1.00} &
  0.25 \\ \cline{4-7} 
\multicolumn{1}{|c|}{} &
  \multicolumn{1}{c|}{} &
  \multicolumn{1}{c|}{} &
  \multicolumn{1}{c|}{$\f^{76/76}$} &
  \multicolumn{1}{c|}{0.00} &
  \multicolumn{1}{c|}{0.00} &
  \textbf{1.00} \\ \hline
\multicolumn{7}{|c|}{\textbf{Setup 2}} \\ \hline
\multicolumn{1}{|c|}{\multirow{2}{*}{$\x^{55}$}} &
  \multicolumn{1}{c|}{\multirow{2}{*}{\begin{tabular}[c]{@{}c@{}}$\f^{55/55}$, \\ $\f^{76/55}$\end{tabular}}} &
  \multicolumn{1}{c|}{\multirow{2}{*}{$\x^{55}$}} &
  \multicolumn{1}{c|}{$\f^{55/55}$} &
  \multicolumn{1}{c|}{0.00} &
  \multicolumn{1}{c|}{0.00} &
  \textbf{1.00} \\ \cline{4-7} 
\multicolumn{1}{|c|}{} &
  \multicolumn{1}{c|}{} &
  \multicolumn{1}{c|}{} &
  \multicolumn{1}{c|}{$\f^{76/55}$} &
  \multicolumn{1}{c|}{0.00} &
  \multicolumn{1}{c|}{0.00} &
  \textbf{1.00} \\ \hline
\multicolumn{1}{|c|}{\multirow{2}{*}{$\x^{76}$}} &
  \multicolumn{1}{c|}{\multirow{2}{*}{\begin{tabular}[c]{@{}c@{}}$\f^{55/76}$, \\ $\f^{76/76}$\end{tabular}}} &
  \multicolumn{1}{c|}{\multirow{2}{*}{$\x^{76}$}} &
  \multicolumn{1}{c|}{$\f^{55/76}$} &
  \multicolumn{1}{c|}{0.00} &
  \multicolumn{1}{c|}{0.00} &
  \textbf{1.00} \\ \cline{4-7} 
\multicolumn{1}{|c|}{} &
  \multicolumn{1}{c|}{} &
  \multicolumn{1}{c|}{} &
  \multicolumn{1}{c|}{$\f^{76/76}$} &
  \multicolumn{1}{c|}{0.00} &
  \multicolumn{1}{c|}{0.00} &
  \textbf{1.00} \\ \hline
\multicolumn{7}{|c|}{\textbf{Setup 3}} \\ \hline
\multicolumn{1}{|c|}{\multirow{4}{*}{\begin{tabular}[c]{@{}c@{}}$\x^{55}$, \\ $\x^{76}$\end{tabular}}} &
  \multicolumn{1}{c|}{\multirow{4}{*}{\begin{tabular}[c]{@{}c@{}}$\f^{55/55}$, \\ $\f^{76/55}$,\\ $\f^{55/76}$, \\ $\f^{76/76}$\end{tabular}}} &
  \multicolumn{1}{c|}{\multirow{2}{*}{$\x^{55}$}} &
  \multicolumn{1}{c|}{$\f^{55/55}$} &
  \multicolumn{1}{c|}{0.00} &
  \multicolumn{1}{c|}{0.00} &
  \textbf{1.00} \\ \cline{4-7} 
\multicolumn{1}{|c|}{} &
  \multicolumn{1}{c|}{} &
  \multicolumn{1}{c|}{} &
  \multicolumn{1}{c|}{$\f^{76/55}$} &
  \multicolumn{1}{c|}{0.00} &
  \multicolumn{1}{c|}{0.00} &
  \textbf{1.00} \\ \cline{3-7} 
\multicolumn{1}{|c|}{} &
  \multicolumn{1}{c|}{} &
  \multicolumn{1}{c|}{\multirow{2}{*}{$\x^{76}$}} &
  \multicolumn{1}{c|}{$\f^{55/76}$} &
  \multicolumn{1}{c|}{0.00} &
  \multicolumn{1}{c|}{0.00} &
  \textbf{1.00} \\ \cline{4-7} 
\multicolumn{1}{|c|}{} &
  \multicolumn{1}{c|}{} &
  \multicolumn{1}{c|}{} &
  \multicolumn{1}{c|}{$\f^{76/76}$} &
  \multicolumn{1}{c|}{0.00} &
  \multicolumn{1}{c|}{0.00} &
  \textbf{1.00} \\ \hline
\end{tabular}
\caption{Mean $P_{miss}$, $P_{fa}$ and AUC over 5 runs on test data for the considered setups. $P_{miss}$ and $P_{fa}$ are found setting the acceptance threshold $\tau=0.5$.}
\label{tab:results}
\end{table}

In \autoref{tab:results} we present the average $P_{miss}$, $P_{fa}$ and $AUC$ score for all settings over 5 runs with different seeds. The first four columns of \autoref{tab:results} define the particular setting, specifying the originals and fakes used during training and testing of the model. $P_{miss}$ and $P_{fa}$ are found for each model by setting an acceptance threshold $\tau=0.5$. The AUC score is found varying the values of $\tau$.

We empirically show the ease of a DL-based model at distinguishing fake CDP in setups 2 and 3. Our approach thus yields perfect performance in the absence of distribution shift at testing time.

Also for setup 1 the system performs perfectly against fakes which distribution was observed during training. However, only in one out of four cases the model could generalize to a distribution shift at the test time. This is the case when the system is trained on the originals and fakes printed with the printer HPI55, but then tested with fakes printed on the printer HPI76.

\begin{figure}[t]
    \centering
    \begin{subfigure}[b]{0.22\textwidth}
        \centering
        \includegraphics[width=\textwidth]{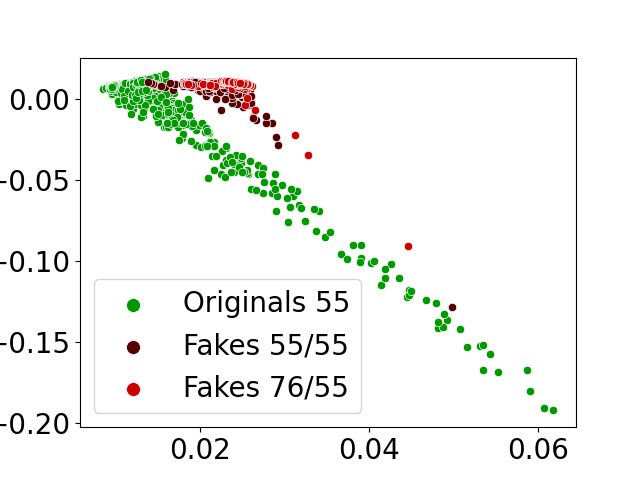}
        \caption{Trained with $\x^{55}$, $\f^{55/55}$}
        \label{fig:pca_x55_f55/55}
    \end{subfigure}
    \hfill
    \begin{subfigure}[b]{0.22\textwidth}
        \centering
        \includegraphics[width=\textwidth]{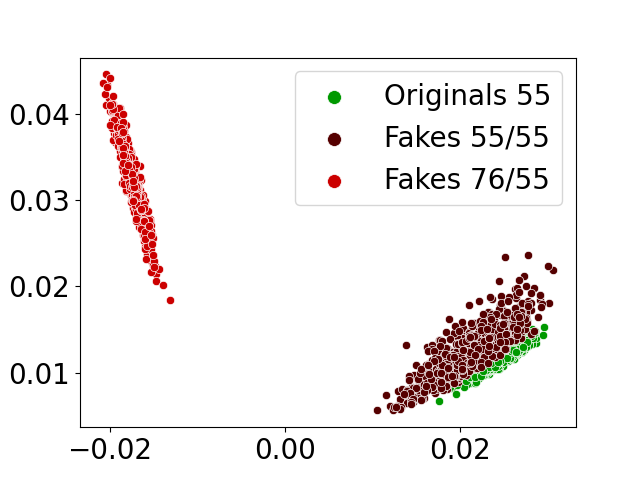}
        \caption{Trained with $\x^{55}$, $\f^{76/55}$}
        \label{fig:pca_x55_f76/55}
    \end{subfigure}
    
    \vfill
    
    \begin{subfigure}[b]{0.22\textwidth}
        \centering
        \includegraphics[width=\textwidth]{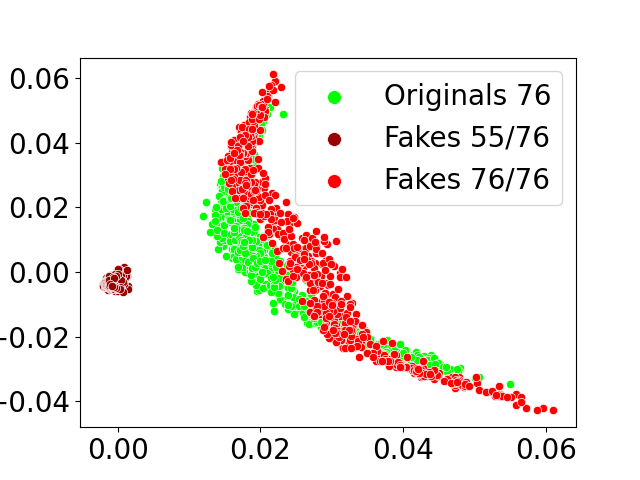}
        \caption{Trained with $\x^{76}$, $\f^{55/76}$}
        \label{fig:pca_x76_f55/76}
    \end{subfigure}
    \hfill
    \begin{subfigure}[b]{0.22\textwidth}
        \centering
        \includegraphics[width=\textwidth]{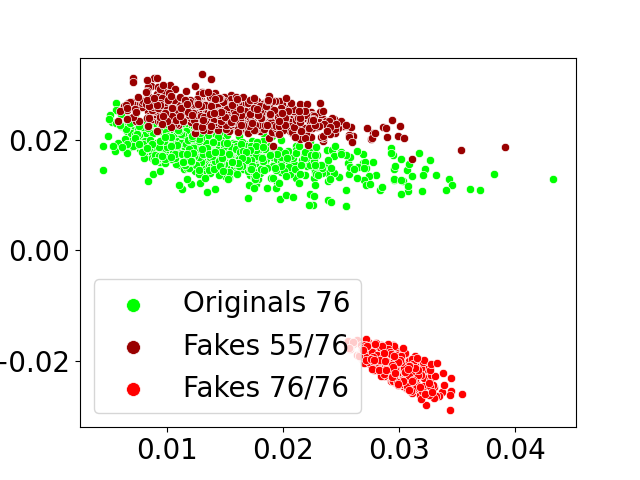}
        \caption{Trained with $\x^{76}$, $\f^{76/76}$}
        \label{fig:pca_x76_f76/76}
    \end{subfigure}
    \caption{2D PCA of ResNet backbone for the four possible configurations in setup 1. Only for case (a) the originals and fakes are almost linearly separable.}
    \label{fig:pca}
\end{figure}

In \autoref{fig:pca} we show the first two dimensions of PCA obtained from features extracted from the ResNet backbone (last layer) for each model in the setup 1.  While in most cases the originals could be distinguished from both sources of fake, only in one case (\autoref{fig:pca_x55_f55/55}) this can be done almost linearly. Thus, the obtained results confirm a known shortcoming of DL methods that require the complete knowledge of fake at the training stage. The alternation of fake' statistics at test time from those assumed at training time might have serious consequences as clearly shown in setup 1. When all fakes are known at training time as considered in settings 2 and 3, the DL authentication system demonstrates an excellent performance.

\section{Conclusion}
\label{sec:conclusion}

In this work, we studied the behaviour of the supervised DL-based authentication system against ML-based fake CDP with 1x1 symbol size. While the system behaves perfectly against the types of fake used at training time, it mostly does not generalize well to fake CDP printed on different printers. In summary, we addressed the following research questions:

\textbf{RQ1: Is it possible to reliably detect ML-based fakes using a two-class classifier?}
The answer is positive. In all cases, our system yields the perfect performance in the absence of a distribution shift.

\textbf{RQ2: Is the supervised classifier robust to a distribution shift of fakes?}
In general, the answer is negative. We show in \autoref{tab:results} that only in one out of four cases the system generalizes to the fakes printed on different printers w.r.t. fake CDP used for training.

While the proposed system can detect fake CDP which distribution was used at training time, it is generally not robust to a distribution shift at test time. The authentication based on the physical references might be a possible solution to this problem. In this respect, in future work, we will investigate the performance of the authentication model based on the physical references, such that information about the statistics of the CDP can be taken into account when authenticating a probe coming from the public domain.

%\vfill
%\pagebreak

\newpage
\bibliographystyle{styles/IEEEbib}
\bibliography{refs}

\end{document}